\documentclass[12pt]{article}
\usepackage{latexsym}
\usepackage{amsmath}
\usepackage{amsfonts}
\usepackage{amssymb}
\usepackage{amscd}
\usepackage{bbm}
\usepackage{fancybox}
\usepackage{cite}
\usepackage{amsmath,amsfonts,amsbsy}
\usepackage{pstricks,pst-node}
\usepackage[small,bf,hang]{caption2}
\usepackage{graphicx}
\usepackage{epsfig}
\usepackage{psfrag}
\usepackage{comment}
\usepackage{mathdots}

\usepackage{float}

\psset{unit=1.3cm,linewidth=.5pt,radius=.2}  

\usepackage{multirow}                     
\usepackage{float}                          
\usepackage{lscape}                         
\usepackage{bm}

\newcommand{\cO}{{\cal O}}

\newcommand{\beq}{\begin{equation}}
\newcommand{\eeq}{\end{equation}}
\newcommand{\bi}{\begin{itemize}}
\newcommand{\ei}{\end{itemize}}
\newcommand{\bt}{\begin{tabular}}
\newcommand{\et}{\end{tabular}}
\newcommand{\bc}{\begin{center}}
\newcommand{\ec}{\end{center}}

\newcommand{\ket}[1]{|#1\rangle}

\newcommand{\bracket}[2]{\langle#1|#2\rangle}
\newcommand{\vev}[1]{\langle#1\rangle}

\newcommand{\be}{\begin{equation}}
\newcommand{\ee}{\end{equation}}
\newcommand{\bea}{\begin{eqnarray}}
\newcommand{\eea}{\end{eqnarray}}
\newcommand{\ba}{\begin{array}}
\newcommand{\ea}{\end{array}}

\def\bbox{{\,\lower0.9pt\vbox{\hrule \hbox{\vrule height 0.2 cm
\hskip 0.2 cm \vrule height 0.2 cm}\hrule}\,}}
\newcommand{\dsl}{\pa \kern-0.5em /}

\newcommand{\rmd}{\mathrm{d}}

\begin{document}

\begin{titlepage}
\begin{center}

\hfill UG-11-56

\vskip 1.5cm

{\large \bf Unitary Truncations and Critical Gravity : a Toy Model  
}

\vskip 1cm

{\bf Eric A.~Bergshoeff$^1$, Sjoerd de Haan$^1$, Wout Merbis$^1$,\\ [.3truecm]
 Massimo Porrati$^2$ and  Jan Rosseel$^1$}

\vskip 25pt

{\em $^1$ \hskip -.1truecm Centre for Theoretical Physics,
University of Groningen, \\ Nijenborgh 4, 9747 AG Groningen, The
Netherlands \vskip 5pt }

{email: {\tt E.A.Bergshoeff@rug.nl, s.de.haan@rug.nl, w.merbis@rug.nl, j.rosseel@rug.nl}} \\
\vskip 15pt

{\em $^2$ \hskip -.1truecm  Center for Cosmology and Particle Physics\\
Department of Physics, New York University\\
New York, NY 10003, USA\vskip 5pt }

{email: {\tt massimo.porrati@nyu.edu}} \\

\end{center}

\vskip 1.5cm

\begin{center} {\bf ABSTRACT}\\[3ex]
\end{center}
We investigate a higher-derivative scalar field model in a fixed $d+1$ dimensional AdS background as a toy model for a gravitational dual to a higher-rank logarithmic CFT. The holographic two-point correlation functions on the boundary agree with higher-rank LCFT correlation functions. For odd rank, the theory allows for a truncation to a nontrivial subspace with non-negative scalar product. We discuss possible implications for  higher-derivative critical gravity theories.

\end{titlepage}

\newpage
%

\section{Introduction}

 Theories of three-dimensional massive higher-derivative gravity~\cite{Deser1982,Bergshoeff:2009hq} have received renewed attention recently~\cite{Li:2008dq,arXiv:0803.3998,arXiv:0805.2610,Giribet2008,arXiv:0903.4573}. For certain ranges of the parameters, these theories have a perturbative spectrum consisting of massive gravitons that are propagated unitarily, making them attractive as toy models for quantum gravity. Many of these theories are not unitary at the non-perturbative level, typically due to the appearance of black holes with negative energy. A notable exception is chiral gravity~\cite{Li:2008dq}, which at least at the classical level admits a truncation to a unitary subspace~\cite{arXiv:0903.4573}. That unitary truncation relies on a split between left and right moving degrees of freedom that is unique to $\mathrm{AdS}_3$.

When considered around an AdS background, massive gravities can lead to an interesting spin-off. For certain so-called `critical' values of the parameters, the massive gravitons disappear from the perturbative spectrum. Such a `critical' gravity theory instead propagates new solutions that were called logarithmic modes. These are characterized by a logarithmic fall-off behavior (in suitable coordinates) towards the AdS boundary, in contrast to the usual massive gravitons that show a power-like  fall-off behavior.
Critical gravities are interesting in the light of the AdS/CFT correspondence \cite{maldacena-1998,Gubser:1998bc,witten-1998}. Indeed, they were conjectured to be dual to logarithmic conformal field theories (LCFTs) \cite{arXiv:0805.2610,Skenderis2009,Grumiller:2009sn,Grumiller:2009mw}. Although typically non-unitary, LCFTs have found applications in condensed matter physics, where they are used in the study of e.g.~critical phenomena, turbulence and percolation. As such, critical gravities might represent gravitational duals of certain strongly coupled condensed matter systems.

One could also try to make sense of critical gravities as toy models for quantum gravity. Then, however, one has to deal with the non-unitarity of these theories. Note that the logarithmic modes, that are responsible for the violation of unitarity, obey {\sl different} boundary conditions than the original massive gravitons. It has been proposed that by imposing strict Brown-Henneaux boundary conditions one could get rid of the problematic logarithmic modes and obtain a theory that is possibly unitary. This approach has been taken recently in two
particular three-dimensional higher-derivative gravity models in AdS: Topologically Massive Gravity (TMG) and New Massive Gravity (NMG). Imposing Brown-Henneaux boundary conditions on critical TMG leads to so-called chiral gravity, that is dual to a two-dimensional chiral CFT \cite{Li:2008dq}. In spite of an apparent non-unitarity at the linear level~\cite{arXiv:0805.2610,Giribet2008} the theory admits a chiral, unitary subsector at the classical level~\cite{arXiv:0903.4573}. TMG however cannot be generalized to higher dimensions; the chiral splitting into right and left movers is unique to a two-dimensional boundary.

New Massive Gravity instead can also be formulated in dimensions higher than three.  At the critical point it becomes a higher-dimensional critical gravity \cite{Lu:2011zk,Deser:2011xc,Alishahiha2011,Bergshoeff:2011ri,arXiv:1104.0674}. Imposing strict Brown-Henneaux boundary conditions leads to a theory that is trivial in the following sense: the full theory describes a massless graviton with zero on-shell energy and its black holes have zero mass and entropy \cite{Lu:2011zk,arXiv:1106.4657,Liu:2009bk}. Modding out these states leaves the vacuum as the only physical state \cite{arXiv:1104.0674}\footnote{For a discussion of truncations away from the critical point see \cite{arXiv:1106.4657,Hyun:2011ej}}.  In \cite{arXiv:1106.4657} it was argued that this triviality of four-dimensional critical gravity is related to a recent proposal by Maldacena \cite{Maldacena2011} that four-dimensional conformal gravity, with specific boundary conditions, is equivalent to Einstein gravity with a cosmological constant.

The aim of imposing specific boundary conditions is to obtain a consistent unitary truncation of the full non-unitary critical theory. On the dual field theory side this means that there should exist a consistent truncation of the LCFT that leads to an ordinary CFT. LCFTs are characterized by the fact that there are fields with degenerate scaling dimensions on which the Hamiltonian acts non-diagonally \cite{Gurarie1993,Flohr2002,Flohr2003}. These degenerate fields form so-called Jordan cells. One of the fields in a Jordan cell corresponds to a zero norm state, while the other fields are referred to as logarithmic partners. The rank of the LCFT then refers to the dimensionality of the Jordan cell. The LCFT dual to critical gravities have rank 2 and operators thus have one logarithmic partner. The truncation mentioned above then corresponds to truncating these logarithmic partners.

In this paper we study holographic scalar LCFTs of rank $r > 2$. The bulk side is made of a scalar field toy model in a fixed AdS background with higher derivatives up to order $2r$. By introducing $r-1$ auxiliary scalar fields, the model can be rewritten as a two-derivative theory. The critical point then corresponds to the point in parameter space where the masses of the $r$ scalar fields degenerate. One can show that $r-1$ higher-order logarithmic modes appear and that these correspond to logarithmic partners of the Klein Gordon scalar mode. For rank $r=2$ the model reduces to the one studied in \cite{Bergshoeff:2011xy}. That case resembles critical NMG in some respects. The theory describes one usual scalar field mode and an associated logarithmic mode, just as in critical NMG. It has been shown that the dual field theory is a LCFT \cite{Kogan:1999bn,Ghezelbash:1998rj}. For a special value of the scaling dimension, these models describe singletons in AdS \cite{Flato1987}.

In this work we show that at the quadratic level, i.e. without introducing interactions, the dual field theory of an odd-rank LCFT allows for a truncation to a unitary CFT. This truncation is different from the one in the rank 2 case mentioned above in the sense that it keeps modes that correspond to the null state {\sl plus} half of the logarithmic modes, whereas in the usual (rank 2) critical gravity proposals the single logarithmic mode is truncated. In section \ref{secScalar} we calculate the 2-point functions of the dual LCFT via holographic methods and show that they indeed agree with the usual form of a rank $r$ LCFT. Next, we show that applying the new truncation leads to the 2-point functions of an ordinary CFT. We also calculate the LCFT scalar product and show that it is positive on the truncated subspace, thus the truncated theory is unitary.
In section \ref{secRank3}, we make the discussion of section \ref{secScalar} explicit for the rank 3 case.
Our conclusions, in particular the implications of our results for critical gravity, can be found in section \ref{discussion}.

\section{A scalar field dual of a rank $r$ LCFT} \label{secScalar}
In this section we propose a scalar field model dual to a rank $r$ LCFT. The model consists of $r$ coupled scalar fields with degenerate masses. Using  holographic methods, we calculate two-point correlation functions on the boundary and we show that these agree with the rank $r$ LCFT two-point correlators. We proceed to calculate the scalar product in the bulk and we point out the existence of a nontrivial subspace for odd rank with positive definite inner product. The corresponding subspace of the higher-rank LCFT is unitary.

The model under consideration is given by the following action:
\begin{equation}\label{Sbulk}
S = - \frac12 \int d^{d+1}x \sqrt{g} \sum\limits_{i,j=1}^{r} \left( A_{ij}\partial_{\mu}\phi_i \partial^{\mu}\phi_j + B_{ij} \phi_i \phi_j \right)\,,
\end{equation}
with the $r$-dimensional matrices given by:
\begin{align}\label{Amatgenr}
A_{ij} & = \left( \begin{array}{cccc}
0		&		\cdots	&		0		&		1\\
\vdots	&				&		1		&		0\\
0		&		\iddots	&				&		\vdots \\
1		&		0		&		\cdots	&		0
\end{array} \right)\,, \qquad
B_{ij}  = \left( \begin{array}{ccccc}
0		&		\ldots	&		0		&		1		&		m^2		\\
\vdots	&				&		1		&		m^2		&		0		\\
0		&		\iddots	&		\iddots	&				&		\vdots	\\
1		&		m^2		&				&				&				\\
m^2		&		0		&		\cdots	&				&		0		\\
\end{array} \right)\,.
\end{align}
The equations of motion take the form:
\begin{align}
(\Box - m^2)\phi_1 & = 0\,, \\
(\Box - m^2)\phi_i & = \phi_{i-1}\,, \qquad {\rm for}\; i=2, \ldots, r\,,
\end{align}
which can be formulated in terms of a single scalar field obeying:
\begin{equation}
(\Box - m^2)^r\phi_r = 0\,.
\end{equation}
The fields $\phi_l$ with $l=1, \ldots, r-1$ can be seen as auxiliary fields used to lower the number of derivatives in the action.

The equations of motion and the on-shell action are invariant under a shift of the scalar fields by:
\begin{equation}\label{shift4}
\phi_i \to \phi_i + \sum\limits_{k=1}^{i-1}\lambda_k \phi_{i-k}
\end{equation}
for general $\lambda_k$ with $1\leq k \leq i-1$. This symmetry is a bulk version of the well-known shift symmetry in logarithmic CFTs.

We will consider this model on a fixed AdS$_{d+1}$ background in the Euclidean Poincar\'e patch:
\begin{equation}
	\rmd s^2 = \frac{1}{z^2}\left( \rmd z^2 + \rmd x_a \rmd x^a \right)\,,
	\label{background}
\end{equation}
with $a=1,\ldots, d$ and the AdS length is set to one. We assume that the scalar field configuration decouples from the metric equations of motions at least up to the order of coefficients that contribute to any divergent terms in the bulk action. This assumption justifies ignoring the back reaction of the scalars on the metric.

\subsection{From $r$ degenerate masses to a rank $r$ LCFT}
We will now calculate the two-point correlation functions on the boundary and show that they form a rank $r$ LCFT. The scalar fields are written in terms of bulk-to-boundary propagators which relate the bulk solution to the pre-specified boundary field $\phi_{i(0)}$:
\begin{align}\label{boundariesgenr1}
\phi_1 (z,x) & = \int d^d x' \left( \phi_{1(0)}(x') G^{\rm KG}(z,x; 0,x')  \right)\,, \\
\phi_i (z,x) & = \int d^d x' \left( \phi_{i(0)}(x') G^{\rm KG}(z,x; 0,x') +  \sum\limits_{j=1}^{i-1} \phi_{j(0)}(x')G^{\rm log^{i-j}}(z,x;0,x')\right)\,, \label{boundariesgenr2}
\end{align}
for $i = 2, \ldots, r$. The functions $G^{\rm KG}$, $G^{\rm log^{k}}$ denote the bulk-to-boundary propagators of the Klein-Gordon mode and the log$^k$ mode respectively. They satisfy the relation:
\begin{align}
(\Box - m^2) G^{\rm KG} & = 0\,, \label{Geom1} \\
(\Box - m^2) G^{\log} & = G^{\rm KG}\,, \label{Geom2}  \\
(\Box - m^2)G^{\log^k} & = G^{\log^{k-1}}\,, \quad {\rm for}\; k=2, \ldots, r-1\,, \label{Geom3}
\end{align}
The bulk-to-boundary propagator $G^{\rm KG}$ can be obtained from solving the homogeneous Klein-Gordon equation in an AdS$_{d+1}$ background. The result is known from \cite{witten-1998} and given by:
\begin{equation}\label{gkg}
G^{\rm KG}(z,x;0,x') = \frac{z^{\Delta}}{(z^2 + |x-x'|^2)^\Delta}\,,
\end{equation}
where $\Delta$, the conformal dimension of the dual operator, is determined by the scalar field mass as the larger root of the equation:
\begin{equation}
\Delta(\Delta -d) = m^2\,.
\end{equation}
To find $G^{\rm log}(z,x;0,x')$ we use the fact that the d'Alembertian does not depend on $m$:
\begin{equation}
[\frac{d}{dm^2},(\Box-m^2)]= -1\,.
\end{equation}
Using this, together with \eqref{Geom1}, we can write $G^{\rm KG}$ as:
\begin{equation}
G^{\rm KG} = -[\frac{d}{dm^2},(\Box-m^2)]G^{\rm KG} = (\Box - m^2) \frac{d}{dm^2} G^{\rm KG}\,.
\end{equation}
Comparing this with \eqref{Geom2} gives:
\begin{equation}\label{glog}
G^{\rm log}(z,x;0,x') = \frac{d}{dm^2} G^{\rm KG} = \frac{1}{2\Delta - d} \log\left( \frac{z}{(z^2 + |x-x'|^2)} \right) \frac{z^{\Delta}}{(z^2 + |x-x'|^2)^{\Delta}} \,.
\end{equation}
The bulk-to-boundary propagator of the higher-order log modes can be obtained by successive application of differentiation with respect to $m^2$:
\begin{align}
G^{\rm log^i}(z,x;0,x') & =  \frac{1}{2^{i-1}} \frac{d^i}{(dm^2)^i} G^{\rm KG} \\
& = \frac{1}{2^{i-1}} \frac{1}{(2\Delta - d)^i} \log^i \left( \frac{z}{(z^2 + |x-x'|^2)} \right) \frac{z^{\Delta}}{(z^2 + |x-x'|^2)^{\Delta}}\\ \nonumber & + \sum_{j=1}^{i-1} \alpha_j \log^j \left( \frac{z}{(z^2 + |x-x'|^2)} \right) \frac{z^{\Delta}}{(z^2 + |x-x'|^2)^{\Delta}}\,,
\end{align}
where $\alpha_j$ are $\Delta$ dependent coefficients. They can be set to zero by adding to $G^{\rm log^i}(z,x;0,x')$ a linear combination of the bulk to boundary propagators $G^{\rm log^j}(z,x;0,x')$, $j<i$.

From the explicit solutions we can calculate one- and two-point correlation functions using AdS/CFT methods.
The bulk action can be written as a surface integral on-shell by integration by parts. At a regulated surface $z=\epsilon$, the on-shell boundary action is:
\begin{equation}\label{sreg}
S = \lim\limits_{\epsilon \rightarrow 0} - \frac12 \int d^dx \sqrt{\gamma} \sum\limits_{i,j=1}^{r} A_{ij} \phi_i (\vec{n}\cdot \vec{\nabla}) \phi_j\,,
\end{equation}
where $A_{ij}$ is given in \eqref{Amatgenr}. The normal derivative is $ (\vec{n}\cdot \vec{\nabla}) = z \partial_z$ and $\sqrt{\gamma} = z^{-d}$, with $\gamma$ the induced metric on the boundary.
This action still contains polynomial and logarithmic divergences in $\epsilon$. These can be eliminated by means of holographic renormalization, as outlined in \cite{Skenderis:2002wp}. Holographic renormalization affects the normalization of the 2-point functions, but it does not change their structure. Since we are primarily interested in the overall structure, we ignore the divergent terms in the action. The proper normalization of the correlation functions may also be obtained using the renormalized result for rank 2, obtained in \cite{Bergshoeff:2011xy} and a set of Ward-type identities relating these to the rank $r$ correlation functions. We refer to the appendix for the details of the normalization process.

In terms of the sources for the boundary operators, i.e. the boundary values of the fields, the on-shell action (\ref{sreg})  can be written as:
\begin{align}\label{Sbound}
& S = \int d^dx d^dx'\Bigg[ \sum\limits_{i=1}^{r} \frac12 \phi_{i(0)}(x)\phi_{r+1-i(0)}(x')\frac{\Delta}{|x-x'|^{2\Delta}} \\ \nonumber
 & + \sum\limits_{k=1}^{r-1} \sum\limits_{l=1}^{r-k} \phi_{l(0)}(x) \phi_{r-k-l+1(0)}(x')
\frac{a_k}{(2\Delta-d)^k}\frac{\Delta}{|x-x'|^{2\Delta}} \times \\\nonumber
&\times \left( \log^k \left( \frac{\epsilon}{|x-x'|^2} \right) + \frac{b_k}{\Delta} \log^{k-1} \left( \frac{\epsilon}{|x-x'|^2} \right) \right) \Bigg]\,,
\end{align}
where $a_k$ and $b_k$ are constants. Following the AdS/CFT logic, we couple the boundary values of the fields to the dual operators as:
\begin{equation}\label{coupling}
\int d^dx \left( \phi_{r(0)} \mathcal{O}^{\rm KG} + \sum\limits_{i=1}^{r-1} c_{r-i} \phi_{i(0)} \mathcal{O}^{\rm log^{r-i}} \right) \,,
\end{equation}
where $c_k$ are constants which may depend on $\Delta$. Then the one-point functions can be determined by functional differentiation with respect to the boundary value of the scalar fields:
\begin{align}
& \frac{\delta S}{\delta \phi_{r(0)}(x)}  =   \vev{\cO^{\rm KG}(x)} = \int d^dx' \phi_{1(0)}(x')\frac{\Delta}{|x-x'|^{2\Delta}}\,, & \\
& \frac{\delta S}{\delta \phi_{r-i(0)}(x)}  =  c_i \vev{\cO^{\rm log^{i}}(x)} = \int d^dx'\Bigg[ \phi_{i+1(0)}(x')\frac{\Delta}{|x-x'|^{2\Delta}} & \\
& +  \sum\limits_{k=1}^{i}  \phi_{i-k+1(0)}(x')
\frac{2 a_k}{(2\Delta-d)^k}\frac{\Delta}{|x-x'|^{2\Delta}} \left( \log^k \left( \frac{\epsilon}{|x-x'|^2} \right) + \frac{b_k}{\Delta} \log^{k-1} \left( \frac{\epsilon}{|x-x'|^2} \right) \right) \Bigg]\,, \nonumber
\end{align}
where now $i = 1, \ldots, r-1$.
After performing the shift symmetry \eqref{shift4}, the one-point functions $\vev{\cO^{\rm log^{i}}(x)}$ can be brought into the following form:
\begin{align}
& c_i \vev{\cO^{\rm log^{i}}(x)} = \int d^dx'\Bigg[ \phi_{i+1(0)}(x')\frac{\Delta}{|x-x'|^{2\Delta}}  \\
& \quad +  \sum\limits_{k=1}^{i}  \phi_{i+1-k(0)}(x')
\frac{\Delta}{|x-x'|^{2\Delta}} \sum\limits_{l=1}^{k} \left(\Lambda_{k-l}  \log^l \left( \frac{\epsilon}{|x-x'|^2} \right) + \Lambda_k \right) \Bigg] \nonumber \,,
\end{align}
where $\Lambda_i$ are constants related to the arbitrary shift parameters $\lambda_i$.
Finally, upon further differentiation with respect to the source, we find that the two-point functions are, up to a normalization factor:
\begin{align}\label{twopoint1}
& \vev{\cO^{\rm KG}(x)\cO^{\rm KG}(x')}  = \vev{\cO^{\rm KG}(x)\cO^{\rm log^n}(x')} = 0, \quad {\rm for}\; n=1,\ldots, r-2\,, \\
& c_{r-1} \vev{\cO^{\rm KG}(x)\cO^{\rm log^{r-1}}(x')}  \sim \frac{\Delta}{|x-x'|^{2\Delta}}\,, \label{twopoint2} \\
& c_i c_j \vev{\cO^{\rm log^i}(x)\cO^{\rm log^j}(x')}  \sim \frac{\Delta}{|x-x'|^{2\Delta}}
\left[  \delta_{m 0} +  \sum\limits_{l=1}^{m} \left(\Lambda_{m-l} \log^l \left(\frac{\epsilon}{|x-x'|^2}\right) + \Lambda_m \right) \right]\,, \label{twopoint3}
\end{align}
with $m = i+j-r+1$ and $\Lambda_m = 0$ for $m \leq 0$. To determine the correct normalization in these two-point correlation functions note that it is sufficient only consider the normalization of the leading order logarithmic term in \eqref{twopoint3}. The freedom of redefining the scalar fields by the shift symmetry \eqref{shift4} can be expressed at the field theory side as the following redefinition of the logarithmic partner operators known in LCFT \cite{Flohr2003}:
\begin{equation}\label{shiftoperator}
\cO^{\log^i} \rightarrow \cO^{\log^i} + \sum\limits_{j=1}^{i} \lambda_j \cO^{\log^{i-j}}\,,
\end{equation}
where we take $\cO^{log^0} = \cO^{\rm KG}$. That this is indeed an allowed redefinition can be seen by performing this shift of the operators in the coupling \eqref{coupling}, together with the shift of the fields \eqref{shift4}. At the level of the two-point correlation functions this invariance allows us to shift the subleading logarithmic terms to our convenience. 

The correct normalization of the two-point functions is derived in the appendix. The result is:
\begin{equation}\label{twopointnorm}
 \vev{\cO^{\log^{i}}(x)\cO^{\log^{j}}(x')} = (2\Delta -d)^{r} \frac{(-2)^m}{m!} \frac{\log^m |x-x'|}{|x-x'|^{2\Delta}} + \textrm{subleading log-terms}\,,
\end{equation}
with $m = i+j-r+1$ and it is valid for $m\geq 0$. Correlation functions with $m<0$ are all null.

\subsection{Comparison with known results}

The two-point correlation functions of a two dimensional rank $r$ LCFT are known from \cite{Flohr2002}. They are given by
\begin{equation}\label{2ptFlohr}
 \vev{\cO^{\rm log^i}(x)\cO^{\rm log^j}(x')} = \frac{1}{|x-x'|^{2\Delta}} \sum\limits_{l=0}^{i+j} D_{i+j-l}\frac{(-2)^l}{l!}\log^l(|x-x'|)\,,
\end{equation}
where the $D$s are constants that satisfy $D_{k} =0$ for $k < r-1$. This implies that the leading order log-term in any two-point function has the power $i+j-r+1$, which agrees with the result given in (\ref{twopointnorm}). The constants $D_k$ can now be related to the scalar field mass $m^2$ by comparing \eqref{2ptFlohr} to \eqref{twopointnorm}. We find that:
\begin{equation}
D_{r-1} = \left(2\Delta - d\right)^r = 2^r \left(1+m^2 \right)^{\frac{r}{2}}\,.
\end{equation}
In the last equality we have used the fact that this only holds in two dimensions, since \eqref{2ptFlohr} is a two dimensional result. This expression is valid for all $r \geq 2$, so we have found the holographic expression for the new anomaly in the rank $r$ LCFT model.

Note that for odd rank $r$, requiring the new anomaly to be real is analogous to requiring that the bulk scalar field satisfies the Breitenlohner-Freedman bound $m^2 \geq -1$ in two dimensions with the AdS length set to one. In fact, when the BF-bound is saturated for all rank $r$ the couplings of the logarithmic partner operators diverge and the new anomaly becomes zero. This was already remarked in the case of $r=2$ in \cite{Kogan:1999bn}.

\subsection{A non-negative scalar product}

The AdS/CFT duality, together with the state/operator duality~\cite{witten-1998},
implies that the normalizable modes of the bulk theory behaving asymptotically
as $\phi_{i+1}\sim z^\Delta\log^i (z)$
are in one-to-one correspondence with the logarithmic CFT states: $O^{\log^i}$.
Therefore, their scalar product must have the same property as the two-point
function in eq.~(\ref{2ptFlohr}); namely, it must be non vanishing whenever
$i+j -r+1\geq 0$.
It is easy to check this property by using the results of ref.~\cite{arXiv:1104.0674},
where a simple method to derive the scalar product of
non-diagonalizable quadratic theories was developed. In our case the method gives a
particularly simple result. Call $\phi^+ (\phi^-)$ the positive (negative) frequency part of the scalar
field $\phi$: $\phi=\phi^+ +\phi^-$; then the scalar product between two normalizable
modes  $\phi^+$, $\psi^+$ is given by:
\begin{equation}{\label{p1}}
\langle \phi^+ | \psi^+ \rangle =
\int d^{d}{\bf x}\, \sqrt{g}g^{00} \sum_{i,j=1}^r\left( A_{ij} \phi^{+*}_i
\stackrel{\leftrightarrow}{\partial_t} \psi^+_j \right)\,,
\end{equation}
where ${\bf x}$  denotes the AdS spatial coordinates.
In the maximally degenerate case, the matrix $A_{ij}$ is given
in equation~(\ref{Amatgenr}). It is nonzero iff $i+j=r+1$. Now, the bulk mode dual to
$O^{\log^i}$ is the normalizable field with asymptotic behavior $z^\Delta \log^i(z)$.
Its nonzero components are the $\phi_l$ with $l\leq i+1$. Likewise, the bulk dual of
$O^{\log^j}$ has nonzero components $\psi_l$ with $ l \leq j+1$. So, the scalar product is
nonzero only if $i+1+j+1\geq r+1$. This is the same condition that holds for the
$D_k$ coefficients in the two-point correlator of a rank-$r$ logarithmic CFT, see
eq.~(\ref{2ptFlohr}).

We now note  that when $r$ is odd, one can define a positive-norm subspace
even in the maximally degenerate theory. It is the subspace defined by
$\phi_i=0$ for $i < n$, where $r=2n-1$, modulo the equivalence relation
$\phi_i \sim \phi_i + \sum_{l=1}^{n-1} \lambda_l \phi_l$.
This is possible because in this subspace the scalar product is
non-negative. In particular all states except $\phi_n$ are null vectors, as is evident from
the scalar product formula:
\begin{equation}{\label{p2}}
\langle \phi^+ | \psi^+ \rangle =
\int d^{d}{\bf x}\, \sqrt{g}g^{00} \phi^{+*}_n
\stackrel{\leftrightarrow}{\partial_t} \psi^+_n .
\end{equation}
When $\Delta$ is in the range $d/2+1 >\Delta > d/2-1$,
an alternative quantization is possible where
the normalizable states behave as $\sim z^{d-\Delta} \log^i(z)$~\cite{klebanov1999}.
The alternative quantization gives similar results to the standard one. The case
$\Delta=d/2-1$ requires a separate analysis, which we shall
discuss in the subsection below.

From the perspective of the AdS/CFT correspondence, the truncation $\phi_i=0$ for $i<n$ amounts to turning off the sources $\phi_{i(0)}$ for $i <n$ in the boundary field theory. Such a truncation can be achieved by imposing that the near boundary behavior of the fields involves powers of $\log z$ of at most $\cO(z^{\Delta} \log^{n-1}z)$. Effectively we are throwing away half of the logarithmic partner modes, while keeping the half which involves lower powers of $\log z$. The boundary action \eqref{Sbound} reduces to:
\begin{equation}\label{Sboundtrunc}
 S[\phi_{n(0)}]  = \Delta \int d^dx d^dx' \left( \frac12 \phi_{n(0)}(x)\phi_{n(0)}(x') \frac{1 }{|x-x'|^{2\Delta}} \right)\,.
\end{equation}
All logarithmic divergent terms have vanished. The boundary correlation functions are now free of logarithmic singularities and they are either proportional to the ordinary CFT two-point function or they vanish. The final result, including the correct normalization is:
\begin{equation}
\vev{\cO^{\rm log^{n-1}}(x)\cO^{\rm log^{n-1}}(x')} = \frac{(2\Delta-d)^r}{|x-x'|^{2\Delta}}\,,
\end{equation}
while all other correlators are zero.

When $r$ is even ($r = 2m$), a similar truncation would not work. If we set to zero all $\phi_i$ for $i \leq m$, the bulk scalar product \eqref{p1}  vanishes. Equivalently, setting the sources $\phi_{i(0)}$ to zero for $i\leq m$ would render the boundary action \eqref{Sbound} trivial and all boundary two point correlators vanish. In the case of rank two, where our model is a spin-0 toy model for critical gravity, this truncation is in essence the one discussed in \cite{arXiv:1106.4657}. Truncating the logarithmic modes in critical gravity by imposing the appropriate boundary conditions will lead to a theory where the surviving massless modes have zero energy and zero norm.

\subsubsection{A special case: the singleton}

When $\Delta=d/2-1$ and $r=2$, the KG mode $\phi\sim z^{d/2-1}$ is a {\em singleton}. When two
modes are identified modulo modes with asymptotic behavior $z^{d/2+1}$, the KG theory
describes only boundary excitations. A ``dipole'' model for singletons was developed
in~\cite{Flato1987}. Setting the AdS radius to $L=1$, the generalization to $d$
dimensions  of the action of ref.~\cite{Flato1987} is:
\begin{equation}\label{p3}
S=\int d^{d+1}x\sqrt{g} \left(\phi_1 [\Box + (d^2/4-1)] \phi_2 - \frac12 \phi^2_1\right) +
S_B.
\end{equation}
Except for the boundary term $S_B$ this is a $r=2$ degenerate scalar with mass
$m^2= -d^2/4+1$, which does not allow for a unitary truncation. The boundary term is where the difference lies.
In the Poincar\'e coordinates of eq.~(\ref{background}) the metric reads $ds^2 = z^{-2}(dz^2 + \eta_{\mu\nu}dx^\mu dx^\nu)$ and the boundary term can be written as:
\begin{equation}\label{p4}
S_B=\lim_{z\rightarrow 0}\int d^dx z^{2-d}
\eta^{\mu\nu}\partial_\mu \phi_2 \partial_\nu \phi_2.
\end{equation}
This boundary term is compatible with the asymptotic behavior:
\begin{eqnarray}\label{p5}
&&\phi_1 \sim 2 A(x) z^{d/2+1},\;\; \\ [.1truecm]\nonumber
&&\phi_2 \sim A(x) z^{d/2+1} \log (z) + B(x)
z^{d/2+1} + z^{d/2-1} \varphi(x), \;\; \ \ \ {\rm with}\ \ \ \partial_\mu \partial^\mu \varphi(x)=0,
\end{eqnarray}
where $A(x),B(x)$ are arbitrary functions of the boundary coordinates.
The boundary term allows for a singleton mode, but this mode must be a
$d$-dimensional free massless scalar. This is one way to see that the
singleton, which saturates the unitarity bound for a $d$-dimensional CFT, must
be a free field.

Besides making room for a singleton, the boundary action $S_B$ also changes the scalar
product. An application of the formulas of~\cite{arXiv:1104.0674} to the action~(\ref{p3})
with boundary term~(\ref{p4}) and boundary conditions~(\ref{p5}) gives:
\begin{equation}\label{p6}
\langle \phi^+ | \phi'^+ \rangle =
\int d^{d}{\bf x}\, \sqrt{g}g^{00} (\phi^{+*}_1
\stackrel{\leftrightarrow}{\partial_t} \phi'^+_2 + 1\leftrightarrow 2 )+
\int d^{d-1} {\rm x}\, \varphi^{+*}({\rm x}) \stackrel{\leftrightarrow}{\partial_t}
\varphi'^+({\rm x}).
\end{equation}
Roman letters denote the spatial coordinates of the $d$-dimensional boundary.
With this definition of the scalar product, the subspace $\phi_1=0$, quotiented by
the identification $\phi_2 \sim \phi_2'$ iff $\varphi=\varphi'$ has a positive
scalar product.

\section{Example: the rank 3 AdS/LCFT scalar model} \label{secRank3}
To illustrate some of the points made in the previous section, it is instructive to take a closer look at the specific case of $r=3$. The bulk field configuration can be written as three coupled scalar fields with degenerate masses or as a single scalar field obeying the sixth-order differential equation

\begin{equation}
	(\Box - m^2)^3 \phi = 0 \,.
\end{equation}
 For two rank 3 fields $\phi$ and $\psi$, the inner product \eqref{p1} becomes:
\begin{align}
\langle \phi^+ | \psi^+ \rangle = &
\int d^{d}{\bf x} \sqrt{g}g^{00} \big\{ (\Box - m^2)^2 \psi^{+\ast} \partial_0 \phi^+ + (\Box - m^2)\psi^{+\ast} \partial_0 (\Box - m^2) \phi^{+}  \\\nonumber
& + \psi^{+\ast} \partial_0 (\Box - m^2)^2 \phi^{+} \big\} \,.
\end{align}
If we decompose the $\phi$ into $\phi = \phi^{\text{KG}} + \phi^{\log} + \phi^{\log^2}$ it can be seen that, among others, the following scalar products hold:
\begin{align}
&	\bracket{\phi^{+\text{KG}}}{\phi^{+{\rm KG}}} =0\,,  \\
&	\bracket{\phi^{+\text{KG}}}{\phi^{+\log^2}} > 0 \,,  \quad
	\bracket{\phi^{+\log^2}}{\phi^{+\log^2}} > 0 \,.
\end{align}
In analogy to \cite{arXiv:1104.0674}, we can consider a state $\ket{\phi^+} = \ket{\phi^{+\log^2}} + \alpha\ket{\phi^{+\text{KG}}} $ such that the norm
\begin{equation}
\bracket{\phi^+}{\phi^+} =  \bracket{\phi^{+\log^2}}{\phi^{+\log^2}} + 2\text{Re}(\alpha \bracket{\phi^{+\text{KG}}}{\phi^{+ \log^2}}) \,,
\end{equation}
can be tuned to be negative. Thus we must conclude that the theory is non-unitary.

There is, however, a non-trivial subspace with positive norm. It is constrained by $\phi^{\log^2} = 0$; i.e. it contains only modes that satisfy the dipole equation $(\Box - m^2)^2 \phi' =0$. The scalar product on this subspace reduces to:
\begin{equation}
\langle \phi^+ | \psi^+ \rangle =
\int d^{d}{\bf x} \sqrt{g}g^{00}  (\Box - m^2)\psi'^{+\ast} \partial_0 (\Box - m^2) \phi'^{+}\,,
\end{equation}
and the only nonzero inner product between the two modes is:
 \begin{align}
	\bracket{\phi^{+\log}}{\phi^{+\log}} > 0 \,.
\end{align}
Thus the norm of the rank 3 theory is positive definite on the rank 2 subspace, which still contains the Klein Gordon and the log modes.
This is to be contrasted with the pure rank 2 case, where the same modes can lead to a negative scalar product. In this case \eqref{p1} gives non-zero values for $\bracket{\phi^{+\log}}{\phi^{+\log}}$ \textit{and} 
$\bracket{\phi^{+\text{KG}}}{\phi^{+\log}}$. Therefore, one can construct a state $\ket{\phi^+} = \ket{\phi^{+\log}} + \alpha\ket{\phi^{+\text{KG}}} $ such that:
\begin{equation}
\bracket{\phi^+}{\phi^+} =  \bracket{\phi^{+\log}}{\phi^{+\log}} + 2\text{Re}(\alpha \bracket{\phi^{+\text{KG}}}{\phi^{+ \log}}) \,,
\end{equation}
can be negative. In the truncated rank 3 case $\bracket{\phi^{+\text{KG}}}{\phi^{+\log}} = 0$, and no negative norm states can be created.

Note that this subspace still contains the Klein Gordon null states that allow for a shift symmetry:
\begin{equation}
\phi' \rightarrow \phi' + \chi\,, \qquad (\Box - m^2)\chi = 0\,.
\end{equation}
From the holographic point of view the truncation $\phi^{\log^2}=0$ amounts to setting the source $\phi_{1(0)}(x)=0$. The remaining scalar fields reduce to:
\begin{align}
\phi_2 (z,x) & = \int d^d x' \phi_{2(0)}(x') G^{\rm KG}(z,x; 0,x') \,,  \\
\phi_3 (z,x) & = \int d^d x' \left( \phi_{3(0)}(x') G^{\rm KG}(z,x; 0,x') + \phi_{2(0)}(x')G^{\rm log}(z,x;0,x') \right)\,,
\end{align}
 where $G^{\rm KG}$ and $G^{\log}$ are given in eqs.~\eqref{gkg} and \eqref{glog}. These are the modes of the rank 2 theory embedded in a
  rank 3  theory whose action is given by  eq.~\eqref{Sbulk} for $r=3$. This embedding makes a non-negative scalar product possible, even though the pure $r=2$ theory does not have a positive definite scalar product. The boundary action written in terms of the field theory sources \eqref{Sbound} simplifies to:
\begin{equation}
 S[\phi_{i(0)}]  = \Delta \int d^dx d^dx' \left( \frac12 \phi_{2(0)}(x)\phi_{2(0)}(x') \frac{1 }{|x-x'|^{2\Delta}} \right)\,.
\end{equation}
The correlation functions of the truncated theory are either null, or proportional to the unitary CFT correlation function:
\begin{align}
\vev{\cO^{\rm KG}(x)\cO^{\rm KG}(x')} & = \vev{\cO^{\rm KG}(x)\cO^{\rm log}(x')} = 0\,, \\
\vev{\cO^{\rm log}(x)\cO^{\rm log}(x')} & = \frac{(2\Delta-d)^3}{|x-x'|^{2\Delta}} \,.
\end{align}
This is to be contrasted with  the correlation functions for the un-truncated rank 3 LCFT model, which are given by:
\vskip .01truecm

\begin{align}
\vev{\cO^{\rm KG}(x)\cO^{\rm KG}(x')} = & \; \vev{\cO^{\rm KG}(x)\cO^{\rm log}(x')} = 0\,, \\[.1truecm]
\vev{\cO^{\rm KG}(x)\cO^{\rm log^2}(x')}  = & \; \vev{\cO^{\rm log}(x)\cO^{\rm log}(x')} = \frac{(2\Delta-d)^3}{|x-x'|^{2\Delta}}\,, \label{cor1} \\[.1truecm]
\vev{\cO^{\rm log}(x)\cO^{\rm log^2}(x')} = & \; \frac{(2\Delta-d)^3}{|x-x'|^{2\Delta}}\left( - 2 \log  |x-x'|  + \Lambda_1 \right)\,, \label{cor2} \\[.1truecm]
\vev{\cO^{\rm log^2}(x)\cO^{\rm log^2}(x')}  = & \; \frac{(2\Delta-d)^3}{|x-x'|^{2\Delta}} \bigg( 2 \log^2 |x-x'| + \Lambda_1 \log  |x-x'| + \Lambda_2 \bigg)\,. \label{cor3} 
\end{align}
Effectively, truncating the theory sets all correlators involving $\cO^{\rm log^2}$ to zero.

\section{Discussion}\label{discussion}
We constructed a free scalar field model in a fixed AdS$_{d+1}$ background which, at the level of two-point correlation functions, is dual to a rank $r$ LCFT. For odd rank the theory has a unitary subspace which can be obtained by truncating half of the logarithmic modes. An easy way to see how this works in the case of rank 3 is to consider the following. Schematically, the two-point correlation functions are:
\begin{equation}
\vev{\cO^i \cO^j} \sim \left( \begin{array}{ccc}
0			&	0			&	{\rm CFT}	\\
0			&	{\rm CFT}	&	{\rm L}	\\
{\rm CFT}	&	{\rm L}	&	{\rm L^2}
\end{array}\right)\,,
\end{equation}
where $i,j = {\rm KG}, \log, \log^2$, CFT represents the CFT two point function \eqref{cor1}, L represents \eqref{cor2} and ${\rm L^2}$ is \eqref{cor3}. When we truncate the theory in the manner described in this article, we throw away all modes which generate the third column and row of this matrix. Hence the only non-zero correlation function is proportional to the ordinary CFT correlation function. The remaining bulk modes have a non-negative scalar product and the truncated theory is unitary. This method can be generalized to arbitrary odd rank, but it fails for even rank LCFTs. By considering the rank 4 case in this manner it is immediately clear why:
\begin{equation}
\vev{\cO^i \cO^j} \sim \left( \begin{array}{cccc}
0			& 	0			&		0		&	{\rm CFT}\\
0			&	0			&	{\rm CFT}	&	{\rm L}\\
0			&	{\rm CFT}	&	{\rm L}		&	{\rm L^2}	\\
{\rm CFT}	&	{\rm L}		&	{\rm L^2}	&	{\rm L^3}	
\end{array}\right)\,.
\end{equation}
Truncating the log$^2$ and the log$^3$ amounts to removing the third and fourth row and column of this matrix. All remaining correlators vanish and the theory only contains null states. Truncating only the log$^3$ mode is insufficient as the remaining theory still contains a rank 2 LCFT. Similar arguments apply for general even rank.

The model described in this paper is a toy model in the sense that it is a non-interacting scalar field model. It remains to be seen whether similar statements can be made for interacting spin-2 models. In order to shed more light on these matters, it could be useful to look at gravitational theories with derivatives up to sixth order, that are described by a Lagrangian of the schematic form:\, \footnote{The case $D=3$ is special since, due to parity violation,
physical modes can be described by first order equations instead of
second-order ones. The same parity violation implies that a
left-moving (right-moving) log mode does not need to pair up with a
right-moving (left-moving) log mode. In particular, there exists a
critical point of generalized massive gravity
\cite{Bergshoeff:2010iy} where the fourth-order theory propagates a
left-moving boundary graviton and a right-moving one that
degenerates with a right-moving log and $\text{log}^2$ mode. The CFT
dual to this critical point has been investigated in
\cite{Grumiller:2010tj} and has been shown to be given by a rank
three LCFT. The two-point functions of this LCFT indeed have the
structure as outlined in this work. We thank Thomas Zojer for
pointing this out to us.} 
\begin{equation} \label{schemform}
{\cal L} \sim \Lambda+ R + R^2 + R\Box R + R^3\,,
\end{equation} 
where $\Lambda$ denotes the cosmological constant and $R$ schematically denotes a curvature tensor or scalar. By suitably adjusting the coefficients of the Lagrangian \eqref{schemform}, these theories can admit a `tri-critical point', where two massive gravitons degenerate with the massless one and where the linearized equations of motion take the form (see \cite{Lu:2011ks} for an example in six dimensions)
\begin{equation} \label{tricriteom}
\mathcal{G}_{\mu \nu}(\mathcal{G}(\mathcal{G}(h))) = 0 \,,
\end{equation}
where $h_{\mu \nu}$, $\mathcal{G}_{\mu \nu}$ denote the perturbation of the metric, resp.\ the linearized Einstein tensor around an AdS background. This linearized Einstein tensor plays a similar role as the Klein-Gordon operator $(\Box - m^2)$ in the spin-0 model discussed in this paper and such a tri-critical gravity theory can be seen as a spin-2 version of the rank 3 scalar field model discussed above. In particular, apart from the usual massless graviton solutions, the equations of motion \eqref{tricriteom} also have solutions obeying log- and $\mathrm{log}^2$-boundary conditions, just as in the scalar field toy model. As in the spin-0 model, the CFT-dual is expected to have the structure of a rank 3 logarithmic CFT. It would be interesting to study the structure of the two-point functions of this logarithmic CFT and to see whether this structure allows for a similar, non-trivial truncation as in the spin-0 toy model.

Note that the above outlined gravity model is fully non-linear and has interactions, that are dictated by gauge invariance. Such gauge-invariant interactions can not be captured by our spin-0 model. A spin-2 version of the toy model discussed in this paper, would thus also be useful to study the fate of the truncations, described in this paper, in a fully interacting model. In the spin-0 case, the truncation to the physical subspace can be described as a truncation to modes that obey the asymptotic boundary conditions:
\begin{equation}\label{p7}
\phi_n\sim z^\Delta \log^{n-1}(z) \mbox{ for } z \rightarrow 0 \qquad r=2n-1.
\end{equation}
The difference between a free theory and an interacting one is that in the latter the boundary conditions~(\ref{p7}) may not be preserved by time evolution. This is reminiscent of the three-dimensional critical Topologically Massive Gravity (TMG) case, where a truncation that only retains modes that obey Brown-Henneaux boundary conditions leads to chiral gravity. In this case, it was argued that the truncation can be rephrased as a truncation to a superselection sector of the full theory, that only includes modes that have zero values for the conserved, left-moving Virasoro charges \cite{arXiv:0903.4573}\footnote{Another possibility is to impose final state boundary conditions on a space-like boundary as in~\cite{Maldacena2011}.}. The consistency of the boundary conditions and their preservation under time evolution is then guaranteed by charge conservation. Whether a similar argument can be made in the context of the above proposed spin-2 model for a truncation along the lines described in this paper, remains an open question.

\subsection*{Acknowledgements}
E.B.~wishes to thank the kind support and hospitality of the CCPP at New York University
where part of this work was initiated.
M.P.~is supported in part by NSF grant PHY-0758032, and by ERC Advanced Investigator Grant n.226455 {\em Supersymmetry, Quantum Gravity and Gauge Fields (Superfields)}. S.d.H, W.M. and J.R. are financed by the Dutch stichting voor Fundamenteel Onderzoek der Materie (FOM).

\appendix
\section{Normalization of the two-point functions}
In section \ref{secScalar} we obtained an expression for the two-point correlation functions \eqref{twopoint1}-\eqref{twopoint3} which is not properly normalized. This appendix is devoted to fixing the normalization and obtaining the relation between the scalar field mass parameter and the normalization found in the LCFT literature. In general, one needs to subtract divergences in the boundary action by a local counterterm action.
 This procedure was done for rank 2 in \cite{Bergshoeff:2011xy}, here we will show how these results can be used to fix the normalization of the rank $r$ two-point correlation functions. 

\subsection{Near-boundary expansion}
From the structure of the bulk-to-boundary propagators \eqref{boundariesgenr2} it is clear that we must relax the usual Fefferman-Graham expansions of the fields to include $\log^k (z)$ type behavior for $k=1, \ldots , i-1$. Allowing for these terms the expansion of the field $\phi_i (z,x)$ in powers of $z$ is of the type:
\begin{align}\label{nearbdy}
\phi_i (z,x) =& \;   z^{d-\Delta} \left( \phi_{i(0)}^{(i)}(x) + \sum\limits_{j=1}^{i-1} \phi_{j(0)}^{(i)}(x) \log^{i-j}(z) + \ldots \right) \\
& \nonumber \qquad + z^{\Delta} \left( \tilde{\phi}_{i(0)}^{(i)}(x) + \sum\limits_{j=1}^{i-1} \tilde{\phi}_{j(0)}^{(i)}(x) \log^{i-j}(z) + \ldots \right) \,,
\end{align}
where the dots represent terms of higher order in $z^2$. We have used the superscript $(i)$ to denote that these are coefficients of the expansion of $\phi_i(z,x)$, while the subscripts $i$, $j$ are related to the power of the $\log(z)$-term they are associated to. 

In the rest of this section, we will show how the equations of motion determine most of the coefficients in this near-boundary expansion  in terms of two sets of undetermined coefficients. In particular, the coefficients $\phi^{(i)}_{j (0)}$ with $j < i \leq r$ can be determined in terms of the coefficients $\phi^{(m)}_{m (0)}$, with $m=1, \cdots, r$. Similarly, the coefficients $\tilde{\phi}^{(i)}_{j (0)}$ with $j < i \leq r$ can be determined in terms of the coefficients $\tilde{\phi}^{(m)}_{m (0)}$, with $m=1, \cdots, r$. The coefficients $\phi^{(m)}_{m (0)}$ are the boundary data and are, according to the AdS/CFT correspondence, identified with the sources for the dual field theory operators. The coefficients $\tilde{\phi}^{(m)}_{m (0)}$ on the other hand correspond to the 1-point functions of the dual field theory operators. They are not determined by the near-boundary analysis. They can however be determined in terms of the boundary data by examining the exact solution of the bulk field equations. We will do this at the end of this section.

Using the metric \eqref{background}, the equations of motion in the Poincar\'e patch of AdS$_{d+1}$ are found as:
\begin{equation}\label{eom}
\left( z^2 \partial_z^2 - (d-1)z\partial_z - m^2 + z^2 \Box_x \right) \phi_{i}(z,x) = \phi_{i-1}(z,x)\,,
\end{equation}
where we have set the AdS length to one and where $\Box_x = \partial^a \partial_a$. 
We can then use these equations of motion to relate the coefficients of the expansion of $\phi_{i}(z,x)$ to the coefficients of the expansion of $\phi_{i-1}(z,x)$. Plugging \eqref{nearbdy} into \eqref{eom} one obtains at order $z^{d-\Delta}$:
\begin{align}\label{coefiter}
& (d-2\Delta)\left((i-1)\phi_{1(0)}^{(i)} \log^{i-2}(z) + \sum\limits_{j=2}^{i-1} (i-j)\phi_{j(0)}^{(i)} \log^{i-j-1}(z) \right)  \\ 
& + \sum\limits_{j=2}^{i-1} (i-j)(i-j+1)\phi_{j-1(0)}^{(i)} \log^{i-j-1}(z)  \nonumber  = \phi_{1(0)}^{(i-1)} \log^{i-2}(z) + \sum\limits_{j=2}^{i-1} \phi_{j(0)}^{(i-1)} \log^{i-j-1}(z) \,.
\end{align}
From this expression we can collect like powers of $\log(z)$ to obtain the following relation between the coefficients $\phi_{k(0)}^{(i)}$ and  $\phi_{k(0)}^{(i-1)}$:
\begin{equation} \label{it1}
(d-2\Delta)(i-k) \phi_{k(0)}^{(i)} + (i-k+1)(i-k)\phi_{k-1(0)}^{(i)} = \phi_{k(0)}^{(i-1)}\,,
\end{equation}
for $k= 2, \ldots, i-1$. For $\phi_{1(0)}^{(i)}$ the relation is:
\begin{equation}
(d-2\Delta)(i-1) \phi_{1(0)}^{(i)} = \phi_{1(0)}^{(i-1)}\,.
\end{equation}
The last equation allows one to solve $\phi_{1(0)}^{(i)}$ recursively in terms of $\phi_{1(0)}^{(1)}$.
\begin{equation} \label{it2}
\phi_{1(0)}^{(i)} = \frac{1}{(i-1)!} \frac{1}{(d-2\Delta)^{i-1}} \phi_{1(0)}^{(1)} \,.
\end{equation}
Using this result and the iterative equation \eqref{it1} for $k=2$ one can then determine $\phi^{(i)}_{2(0)}$ in terms of $\phi^{(1)}_{1(0)}$ and $\phi_{2(0)}^{(2)}$. Continuing in this manner all the coefficients $\phi^{(i)}_{j (0)}$, with $j < i$, in the expansion of $\phi_i (z,x)$ can be determined in terms of $\phi_{k(0)}^{(k)}$, with $1 \leq k < i$. So for every $i$, the expansion of $\phi_i(z,x)$ introduces a new undetermined leading order coefficient $\phi_{i(0)}^{(i)}$. Since there are $r$ scalar fields in total, there are also $r$ independent coefficients, namely the $\phi_{m(0)}^{(m)}$ for $m = 1, \ldots, r$.

At order $z^{\Delta}$ we can derive a similar recursion relation between the coefficients $\tilde{\phi}_{k(0)}^{(i)}$ and $\tilde{\phi}_{k(0)}^{(i-1)}$. 
\begin{equation}\label{phitildeiter}
(2\Delta-d)(i-k) \tilde{\phi}_{k(0)}^{(i)} + (i-k+1)(i-k)\tilde{\phi}_{k-1(0)}^{(i)} = \tilde{\phi}_{k(0)}^{(i-1)}\,,
\end{equation}
for $k= 2, \ldots, i-1$, while for $\tilde{\phi}_{1(0)}^{(i)}$ the relation is:
\begin{equation}
(2\Delta-d)(i-1) \tilde{\phi}_{1(0)}^{(i)} = \tilde{\phi}_{1(0)}^{(i-1)}\,.
\end{equation}
Again, for $r$ scalar fields, there are $r$ independent coefficients $\tilde{\phi}_{i(0)}^{(i)}$, with $i=1,\ldots, r$. These are the precisely the coefficients of order $z^{\Delta}$ (so with no $\log(z)$ behavior). In the following, we will drop the superscript $(i)$ from the $\tilde{\phi}^{(i)}_{i(0)}$ (but not yet from the $\phi^{(i)}_{i(0)}$) and we will simply denote these coefficients as $\tilde{\phi}_{i(0)}$. 

In order to calculate two-point correlation functions we need to determine $\tilde{\phi}_{i(0)}$ as a function of the sources $\phi_{i(0)}^{(i)}$. This can be done by examining the exact solutions given in eqs.\ \eqref{boundariesgenr1}-\eqref{boundariesgenr2} for small $z$. Let us for convenience repeat the solution, including the superscript $(i)$:
\begin{equation}
\phi^{(i)}_i (z,x)  = \int d^d x' \left( \phi^{(i)}_{i(0)}(x') G^{\rm KG}(z,x; 0,x') +  \sum\limits_{j=1}^{i-1} \phi^{(i)}_{j(0)}(x')G^{\rm log^{i-j}}(z,x;0,x')\right)\,,
\end{equation}
where the Green's functions $G^\mathrm{KG}$, $G^{\log^i}$ are given in \eqref{gkg}, \eqref{glog}. Note that the boundary data (boundary values) of the fields $\phi^{(i)}_i (z,x)$ can be read off from this expression as the coefficients $\phi^{(i)}_{i (0)}$. It is thus in terms of these $\phi^{(i)}_{i (0)}$ that we will try to express $\tilde{\phi}_{i(0)}$. Near the boundary (for small $z$), the contributions of order $z^{\Delta}$ are given by:
\begin{equation}\label{expand}
\int d^d x' \frac{1}{|x-x'|^{2\Delta}}\left\{ \phi_{i(0)}^{(i)}(x')  + \sum\limits_{j=1}^{i-1} \phi_{j(0)}^{(i)}(x') \frac{2}{(2(2\Delta-d))^{i-j}} \log^{i-j} \left( \frac{z}{|x-x'|^{2}}\right)  \right\}\,.
\end{equation}
As outlined in the previous section, the independent coefficients $\tilde{\phi}_{i(0)}$ are the components in \eqref{expand} that have no $\log(z)$ behavior:
\begin{equation}\label{phitilde}
\tilde{\phi}_{i(0)}(x) = \int d^d x' \left\{ \frac{\phi_{i(0)}^{(i)}(x')}{|x-x'|^{2\Delta}}  + \sum\limits_{j=1}^{i-1} \phi_{j(0)}^{(i)}(x') \frac{2}{(2(2\Delta-d))^{i-j}} \frac{\left(-2\log|x-x'|\right)^{i-j}}{|x-x'|^{2\Delta}} \right\} \,.
\end{equation}
The sources which will be coupled to the field theory operators are the boundary data $\phi_{m(0)}^{(m)}$, with $m=1,\ldots, r$, and we need to find an expression for $\tilde{\phi}_{i(0)}$ in terms of these. To do this we note that, as we are interested in  the normalization of the two-point functions,  we are only interested in the leading logarithmic order of the two-point correlation functions. This is because the subleading contributions may be shifted away by the invariance:
\begin{equation}\label{shift}
\phi_i \to \phi_i + \sum\limits_{k=1}^{i-1}\lambda_k \phi_{i-k}\,.
\end{equation}
 In order to obtain the right relation for $\tilde{\phi}_{i(0)}$ we may also obtain an expression for it using the fact that $\tilde{\phi}_{m(0)}^{(i)}$ is the $\log^{i-m} z$ component of \eqref{expand}. Using this and \eqref{phitildeiter} and keeping only the leading logarithmic terms for every $\phi_{j(0)}^{(k)}$, we arrive at the following expression for $\tilde{\phi}_{i(0)}$ as a function of the boundary fields $\phi_{i(0)}^{(i)}$:
\begin{equation}\label{phitilde2}
\tilde{\phi}_{i(0)}(x) = \int d^d x' \frac{1}{|x-x'|^{2\Delta}}\left\{ \sum\limits_{j=1}^{i} \phi_{j(0)}^{(j)}(x') \frac{2}{(i-j)!} \frac{\left(-2\log|x-x'|\right)^{i-j}}{(2\Delta-d)^{i-j}} \right\}\,.
\end{equation}
Now that everything is defined in terms of the fields $\phi_{i(0)}^{(i)}$, we may drop the superscript ${(i)}$ to ease the notation. From the above expression the following identity may be derived.
\begin{equation}\label{dphiid}
\frac{\partial \tilde{\phi}_{i(0)}}{\partial \phi_{j(0)}} = \frac{1}{(i-j)} \frac{d}{dm^2} \frac{\partial \tilde{\phi}_{i(0)}}{\partial \phi_{j+1(0)}} + {\rm subleading \; log \; terms}\,, \quad {\rm for}\; j\leq i-1\,.
\end{equation}
This relation will prove to be useful later on.

\subsection{Normalization of the holographic 2-pt correlation functions}
We are now ready to consider the boundary two-point correlation functions of this theory. The sources are coupled to the dual operators according to \eqref{coupling} which we repeat here for convenience:
\begin{equation}\label{coupling2}
\int d^dx \left( \phi_{r(0)} \mathcal{O}^{\rm KG} + \sum\limits_{i=1}^{r-1} c_{r-i} \phi_{i(0)} \mathcal{O}^{\rm log^{r-i}} \right) \,.
\end{equation}
Here $\mathcal{O}^{\rm log^i}$ is the operator corresponding to the $i$-th logarithmic partner operator of $\mathcal{O}^{\rm KG}$. The $c_i$ are some normalization constants which may depend on $\Delta$.

Following the holographic dictionary the two-point correlation functions are obtained by functional differentiation of the on-shell renormalized action with respect to the sources:
\begin{equation}\label{2ptholdict}
c_i c_j \vev{\cO^{\rm log^i}(x) \cO^{\rm log^j}(x')} = - \frac{\delta^2 S_{\rm ren}[\phi_{k(0)}] }{\delta \phi_{r-i(0)}(x)\delta \phi_{r-j(0)}(x')}\,.
\end{equation}
This requires knowledge of the renormalized on-shell action as a function of the sources $\phi_{k(0)}$ for $k=1,\ldots,r-1$. Fortunately the renormalized one and two-point correlation functions for $r=2$ have already been calculated in \cite{Bergshoeff:2011xy}. They are:
\begin{align}
\vev{\cO^{\rm KG}(x)}_{\rm ren}^{r=2} & = (d-2\Delta)\tilde{\phi}_{1(0)}(x) \,, \label{1ptKGr2} \\
c_1 \vev{\cO^{\rm log}(x)}_{\rm ren}^{r=2} & = (d-2\Delta)\tilde{\phi}_{2(0)}(x) \,, \label{1ptlogr2} \\
\vev{\cO^{\rm KG}(x) \cO^{\rm KG}(x')}_{\rm ren}^{r=2} & = 0 \label{2ptKGKGr2} \,, \\
c_1 \vev{\cO^{\rm KG}(x) \cO^{\rm log}(x')}_{\rm ren}^{r=2} & = \frac{(2\Delta - d)}{|x-x'|^{2\Delta}}  \,,\label{2ptKGlogr2} \\
c_1^2 \vev{\cO^{\rm log}(x) \cO^{\rm log}(x')}_{\rm ren}^{r=2} & = \frac{1}{|x-x'|^{2\Delta}} \left(-2 \log|x-x'| + \lambda \right)\,, \label{2ptloglogr2} 
\end{align}
where $\tilde{\phi}_{i(0)}(x)$ with $i=1,2$ are defined as in \eqref{phitilde}.

The standard normalization for the two-point functions of a rank 2 LCFT  (in two dimensions) is:
\begin{align}
\vev{\cO^{\rm KG}(x) \cO^{\rm KG}(x')}_{\rm ren}^{r=2} & = 0 \,,\\
\vev{\cO^{\rm KG}(x) \cO^{\rm log}(x')}_{\rm ren}^{r=2} & = \frac{D_1}{|x-x'|^{2\Delta}} \,,\\
\vev{\cO^{\rm log}(x) \cO^{\rm log}(x')}_{\rm ren}^{r=2} & = \frac{1}{|x-x'|^{2\Delta}} \left(-2 D_1 \log|x-x'| + \lambda \right) \,.
\end{align}
This can be matched with the holographic result by taking $c_1 = 1/(2\Delta -d)$. Then the new anomaly $D_1$ in two dimensions can be expressed in terms of the bulk scalar field mass as:
\begin{equation}
D_1 = (2\Delta -d)^2 = (d^2 + 4m^2) = 4(1+m^2) \,.
\end{equation}
The constant $\lambda$ can be rescaled by shifting the operators $\cO^{\log} \rightarrow \cO^{\log} + \lambda' \cO^{\rm KG}$. This corresponds holographically to the reparameterization invariance of the bulk scalar fields $\phi_2 \rightarrow \phi_2 + \lambda'\phi_1$, as can be seen from the coupling of the operators to the sources in \eqref{coupling2}. Thus the only normalization which needs to be fixed holographically to establish a well defined dictionary is the value of the new anomaly.

The way we defined the $\tilde{\phi}_{i(0)}$ in \eqref{phitilde} is such that for any rank $r$, $\tilde{\phi}_{i(0)}$ is the same. The only difference is that the rank $r$ theory will have more $\tilde{\phi}_{i(0)}$ then any rank $r'$ theory with $r>r'$. But since the renormalized one-point functions in \eqref{1ptKGr2} and \eqref{1ptlogr2} only depend on $\tilde{\phi}_{1(0)}$ and $\tilde{\phi}_{2(0)}$, the result carries over to general rank $r$.
\begin{align}
\vev{\cO^{\rm KG}(x)}_{\rm ren}^{r=2} & = \vev{\cO^{\rm KG}(x)}_{\rm ren}^{r}  =   (d-2\Delta)\tilde{\phi}_{1(0)}(x) \label{1ptKGr} \,,\\
c_1 \vev{\cO^{\rm log}(x)}_{\rm ren}^{r=2} & = c_1 \vev{\cO^{\rm log}(x)}_{\rm ren}^{r}  =  (d-2\Delta)\tilde{\phi}_{2(0)}(x) \,. \label{1ptlogr} 
\end{align}
This leads, by use of \eqref{2ptholdict}, to the renormalized two point function for general rank $r$:
\begin{equation}
c_{r-1} \vev{\cO^{\rm KG}(x) \cO^{\rm log^{r-1}}(x')}_{\rm ren}^{r}  = \frac{(2\Delta - d)}{|x-x'|^{2\Delta}}\,. \label{2ptKGlogr}
\end{equation}
Then by using the identity \eqref{dphiid} and by assuming that the one-point correlation functions are all linear in $\tilde{\phi}_{i(0)}$, i.e. assuming that the scalar field theory is free and that there are no interactions or contact terms, one can derive a holographic Ward-type identity:
\begin{align}\label{Ward1}
c_{i+1}c_j \vev{\cO^{\log^{i+1}}(x)\cO^{\log^j}(x')}  = &  \frac{1}{i+j-r+2} \frac{d}{dm^2}\left(c_i c_j \vev{\cO^{\log^{i}}(x)\cO^{\log^j}(x')}\right)  \\  \nonumber & + O\left(\frac{\log^{i+j-r+1}|x-x'|}{|x-x'|^{2\Delta}}\right)\,.
\end{align}
Repeating this procedure while starting from $\vev{\cO^{\rm KG}(x)\cO^{\log^{r-1}}(x')}$ one can write down the Ward identity:
\begin{equation}\label{Ward3}
c_{i}c_{r-1} \vev{\cO^{\log^{i}}\cO^{\log^{r-1}}} = \frac{1}{i!} \left(\frac{d}{dm^2}\right)^{i} \left(  c_{r-1} \vev{\cO^{\rm KG}\cO^{\log^{r-1}}} \right) + O\left(\frac{\log^{i-1}|x-x'|}{|x-x'|^{2\Delta}}\right)\,.
\end{equation}
This is valid for all $i = 1,\ldots, r-1$. From \eqref{2ptKGlogr} it is apparent that differentiation with respect to $m^2$ of the two-point function will introduce $\log|x-x'|$ terms. Repeating the differentiation will increase the power of these terms one step at the time. The leading logarithmic term on the right hand side of \eqref{Ward3} will thus be of order $\log^i|x-x'|$. 

By applying the shift invariance \eqref{shiftoperator} to \eqref{Ward3} we learn that we may always add terms of order $\log^{i-1}|x-x'|$ to the correlation function. So the only term for which we need to fix the normalization is the $\log^i|x-x'|$ term. This justifies restricting to the leading order logarithmic terms earlier. Using \eqref{Ward3} and \eqref{2ptKGlogr} we find:
\begin{equation}\label{2ptlogilogr}
c_{i}c_{r-1} \vev{\cO^{\log^{i}}(x)\cO^{\log^{r-1}}(x')} = \frac{1}{(2\Delta -d)^{i-1}} \frac{(-2)^i}{i!} \frac{\log^i |x-x'|}{|x-x'|^{2\Delta}} + O\left(\frac{\log^{i-1}|x-x'|}{|x-x'|^{2\Delta}}\right)\,.
\end{equation}
Once the normalization for these correlators is fixed, it is simultaneously fixed for all other correlators. This can be seen from the two-point functions given in the paper (equation (28)), from which one may derive that $\vev{\cO^{\log^i}\cO^{\log^{r-1}}} = \vev{\cO^{\log^{i+k}}\cO^{\log^{r-1-k}}}$.

The above result may now be compared with the two-point correlation functions of the two-dimensional rank $r$ LCFT given in \cite{Flohr2002}:
\begin{equation}\label{2ptFlohr2}
 \vev{\cO^{\rm log^i}(x)\cO^{\rm log^j}(x')} = \frac{1}{|x-x'|^{2\Delta}} \sum\limits_{l=0}^{i+j} D_{i+j-l}\frac{(-2)^l}{l!}\log^l(|x-x'|)\,,
\end{equation}
with $D_k = 0$ for $k<r-1$. The constants $D_k$ are determined holographically in terms of the scalar field mass, which is the only free parameter on the gravity side. 

Comparing \eqref{2ptKGlogr} to \eqref{2ptFlohr2} with $i=0$ and $j=r-1$ we find that:
\begin{equation}\label{Dr1}
D_{r-1} = \frac{(2\Delta-d)}{c_{r-1}}\,.
\end{equation}
If we consider the correlation function with the highest order log operators possible, according to \eqref{2ptlogilogr} we have:
\begin{equation}
\label{2ptlogrlogr}
c^2_{r-1} \vev{\cO^{\log^{r-1}}(x)\cO^{\log^{r-1}}(x')} = \frac{1}{(2\Delta -d)^{r-2}} \frac{(-2)^{r-1}}{(r-1)!} \frac{\log^{r-1} |x-x'|}{|x-x'|^{2\Delta}} + {\rm subleading}\,.
\end{equation}
Comparing this to \eqref{2ptFlohr2}, now with $i=j=r-1$ we have:
\begin{equation}\label{Dr2}
D_{r-1} \frac{(-2)^{r-1}}{(r-1)!} = \frac{1}{(2\Delta -d)^{r-2}} \frac{(-2)^{r-1}}{(r-1)!} \frac{1}{c_{r-1}^2}\,.
\end{equation}
From \eqref{Dr1} and \eqref{Dr2} we may find an expression for $D_{r-1}$ in terms of the scalar field mass:
\begin{equation}
D_{r-1} = (2\Delta -d)^r = 2^r (1+m^2)^{\frac{r}{2}}\,.
\end{equation}
This ends the comparison with the two-point correlation functions in two-dimensional rank $r$ LCFTs.

\providecommand{\href}[2]{#2}\begingroup\raggedright\endgroup

\end{document}